# Ising Dirac fermions across a topological phase transition

Aoqian Zhang[1†], Yaqi Ma[1†], Yifei Jin[1†], Nan Zhang[1†], Wentao Jiang[1], Tianyu Qiao[1], Ivana Wang[1], Ulf, Lampe[1], Kenji Watanabe[2], Takashi Taniguchi[2], Tze Kin Cheung[1], Junwei Liu[1,3], Shilin Huang[1], Xi Dai[1], Hoi Chun Po[1,3], Ning Wang[1,4*] and Kaifei Kang[1,3*]

[1]Department of Physics, The Hong Kong University of Science and Technology, Hong Kong SAR, China

[2]National Institute for Materials Science, 1-1 Namiki, 305-0044 Tsukuba, Japan

[3]IAS Center for Quantum Matter, Hong Kong University of Science and Technology, Hong Kong SAR, China

[4]State Key Laboratory of Optical Quantum Materials, Hong Kong University of Science and Technology, Hong Kong SAR, China

[†]These authors contributed equally: Aoqian Zhang, Yaqi Ma, Yifei Jin and Nan, Zhang

[*]Emails: phwang@ust.hk, kfkang@ust.hk

**Dirac fermions have attracted significant interest due to their relativistic dispersions and close connections to topological physics, yet they are generally expected to be gapped in two-dimensional systems with strong Ising spin–orbit coupling, making their realization in such materials an outstanding challenge. Here we report the emergence of six-fold degenerate Dirac fermions in an Ising moiré system across a quantum spin Hall transition in twisted $WSe_2$. In a 3.65° device, we observe a quantum spin Hall phase at high electric fields with nearly quantized resistance $\frac{h}{2e^2}$, and a Dirac semimetal phase over a broad range of electric fields near zero field. Magnetotransport measurements of the Dirac phase exhibit a half-integer Landau fan sequence, characteristic of Dirac fermions, with six-fold degeneracy on the hole-doped side and two-fold degeneracy on the electron-doped side. Temperature dependence shows weakly metallic behavior consistent with a semimetallic state. Our twist-angle-dependent transport measurements map out a complete phase diagram and identify a critical twist angle of θ ≈ 3.3°, establishing the phase boundary between the quantum spin Hall and Dirac semimetal regimes. Our work establishes a new route to realizing Dirac fermions in strongly spin–orbit-coupled moiré systems through a topological phase transition, providing a promising platform for high-mobility spintronics.**

## Main

Dirac fermions are quasiparticles characterized by linearly dispersing energy bands and a π Berry phase, whose unusual electronic properties and deep connection to topological physics have attracted sustained interest over the past two decades[1-6]. In two-dimensional systems, spin-isotropic Dirac fermions were first observed in graphene[7-11] and have since been extensively studied in a variety of material platforms using complementary experimental probes[12-20]. In transport measurements, half-integer Landau level quantization provides a hallmark signature of Dirac fermions and their associated π Berry phase[7-9,21]. However, in systems with strong Ising spin–orbit coupling (SOC), Dirac points are generally expected to acquire a mass gap, driving the system into either a topological or trivial insulating state[22,23]. Although theoretical studies have proposed that Dirac fermions may survive under specific nonsymmorphic crystal symmetries or fine-tuned conditions[24-28], experimental realizations of Dirac quasiparticles robust against strong spin–orbit interactions remain elusive.

An alternative route to realizing Dirac fermions is through band inversion and gap closing at topological phase transitions, where symmetry-enforced band crossings can emerge at phase boundaries[29,30](Fig. 1a). This approach offers a general strategy to access relativistic quasiparticles even in strongly spin–orbit-coupled systems. Twisted transition metal dichalcogenides (TMDs) are a particularly promising platform for this realization[31,32]. In these systems, moiré bands arise from K and K' valley electrons of monolayer semiconductors, where strong Ising SOC locks spins to the out-of-plane direction[33], producing spin - valley - locked electronic states distinct from those in isotropic systems such as graphene.

Recent studies have demonstrated tunable quantum anomalous Hall effects and quantum spin Hall (QSH) phases in twisted TMDs, establishing moiré materials as a highly controllable platform for topology[34-43]. Across the topological phase transition, the topological gap is expected to close with linear crossing, potentially hosting Dirac excitations[29,30]. These quasiparticles inherit strong out-of-plane spin polarization from Ising SOC, forming a new class of Ising Dirac fermions. In contrast to graphene, where Dirac fermions arise from spin-degenerate isotropic bands, these states are intrinsically spin–valley locked and strongly spin-polarized. Moreover, because inversion symmetry is explicitly broken in the triangular lattice of monolayer $WSe_2$ and its moiré superlattice, the Dirac gap closing can extend over a finite parameter range rather than being confined to a single fine-tuned point[30] (Fig. 1b). This opens the possibility of experimentally stabilizing and probing Dirac fermions in a broad region of parameter space, making twisted $WSe_2$ a promising platform for studying Dirac physics at topological phase transitions.

Here we report the experimental realization of six-fold degenerate Dirac fermions in twisted $WSe_2$, emerging across a quantum spin Hall (QSH) phase transition under strong Ising spin–orbit coupling. In devices with a twist angle of 3.65°, we observe a QSH phase at high electric fields with nearly quantized resistance of $\frac{h}{2e^2}$, and a Dirac semimetal phase over a broad range of electric fields near zero field. Magnetotransport measurements reveal a half-integer Landau fan sequence, characteristic of Dirac fermions with a π Berry phase, exhibiting six-fold degeneracy on the hole-doped side and two-fold degeneracy on the electron-doped side. Systematic twist-angle-dependent transport measurements map out a complete phase diagram and identify a critical twist angle of approximately 3.3°, marking the boundary between the QSH and Dirac semimetal regimes. These results demonstrate that Dirac fermions can persist under strong Ising spin–orbit coupling and establish twisted transition metal dichalcogenides as a versatile platform for exploring spin-polarized Dirac physics, topological phase transitions, and high-mobility spintronic functionalities.

**Electrical characterization of the device at $B_{\perp} = 0\ T$**

We first characterize the electrical transport of a 3.65° twisted $WSe_2$ device at $B_{\perp} = 0$ T and a base temperature of $T = 1.5$ K. Figures 1d and 1e show the longitudinal resistance $R_{xx}$ and nonlocal resistance $R_{nl}$ as functions of moiré filling factor $v$ and vertical electric field $E$. The inset illustrates the measurement configurations, respectively. In both measurements, the device is biased with a low-frequency harmonic current (< 10 nA, < 37 Hz) between $I^{+}$ and $I^{-}$, and the voltage drop is measured between $V^{+}$ and $V^{-}$.

For the local measurement (Fig. 1d), we observe pronounced $R_{xx}$ peaks at commensurate moiré filling factors of $v = -1$ and -2. At $v = -1$, $R_{xx}$ exceeds 1 MΩ, consistent with a trivial Mott insulator at half band filling. At $v = -2$, weak resistance peaks appear only around $|E| \approx 0.5\ V \cdot nm^{-1}$, accompanied by a pronounced nonlocal response in the same electric field range (Fig.1e). The linecut of $R_{xx}$ along electric field at fixed $v = -2$ reveal a plateaued resistance $\sim \frac{h}{2e^2}$ (Fig. 1f bottom panel). The observed nearly quantized $R_{xx}$ together with pronounced $R_{nl}$ is consistent with a quantum spin Hall phase with a single pair of helical edge states in the bulk gap. The topological protection of the helical edge conduction is further supported by a pronounced negative in-plane magnetoconductance at the same electric field range (Extended Data Fig. 1).

At smaller electric fields of $|E| < 0.3\ V \cdot nm^{-1}$, $R_{xx}$ is significantly reduced to < 2 kΩ (Fig. 1f ). Both the nonlocal signal and negative in-plane magnetoconductance vanish in this regime, indicating bulk-dominated transport distinct from the QSH phase (Fig. 1e, Extended Data Fig. 1).

**Landau Fan at $E = 0$ V/nm in a 3.65° twisted $WSe_2$ device**

Next, we investigate magnetotransport of the weak resistance state near $E = 0\ V \cdot nm^{-1}$ at $v = -2$. Figures 2a and 2b show the longitudinal ($R_{xx}$) and Hall ($R_{xy}$) resistances as functions of moiré filling factor $v$ and out-of-plane magnetic field $B_{\perp}$. Above $B_{\perp} \sim$ 4T, pronounced minima in $R_{xx}$ emerge near $v = -2$.

On the hole-doped side ($v < -2$), the positions of $R_{xx}$ minima evolve linearly with $B_{\perp}$, consistent with the formation of Landau levels (LL). The slope of each LL follows the Středa formula

$$\frac{\partial \nu}{\partial B_{\perp}} = \frac{g_{LL}\nu_{LL}}{n_M}\frac{e}{h},$$

Where $g_{LL}$ is the degeneracy, $\nu_{LL}$ is the Landau level index and $n_{\mathrm{M}}$ is the moiré density. The extracted $g_{LL}\nu_{LL}$ values follow the sequence -3, -6, -9, -15, -21, -27 and -33. Excluding -6, the sequence exhibits a regular degeneracy of $g_{LL}$ = 6. The anomalous $g_{LL}\nu_{LL}$ = -6 likely arises from magnetic-field-induced splitting of the second LL, enhanced by electron–electron interactions at lower carrier density (Ref. [44-46]). At higher densities, screening suppresses interaction effects and the LLs remain six-fold degenerate.

On the electron-doped side ($\nu > -2$), the lowest LL follows a linear trajectory with slope corresponding to $g_{LL}\nu_{LL} = 1$, while higher LLs evolve into arch-shaped features, with corresponding plateaus in $R_{xy}$, complicating direct application of the Středa analysis. To resolve the LL indexing, we take magnetic-field linecuts at fixed filling, as shown in Fig. 2c. At $\nu = -1.57$, well-developed quantum Hall plateaus appear at $\frac{h}{3e^2}$, $\frac{h}{5e^2}$, $\frac{h}{7e^2}$, $\frac{h}{9e^2}$, $\frac{h}{11e^2}$, and $\frac{h}{13e^2}$, accompanied by vanishing $R_{xx} < 0.01\frac{h}{e^2}$. These correspond to $g_{LL}\nu_{LL}$ = 3, 5, 7, 9, 11, and 13, together with the lowest level at $g_{LL}\nu_{LL} = 1$, forming a sequence 1, 3, 5, 7, 9, 11 and 13 with $g_{LL} = 2$. Similar electron-side sequences and Landau fan features are observed at other moiré fillings in the same device (Extended Data Fig. 2) and in an additional device with twist angle 3.56° (Extended Data Fig. 3), demonstrating the robustness of these observations.

Figure 2d summarizes the hole-like (h-LL), electron-like (e-LL) and symmetry-breaking (SB-LL) trajectories, highlighted in red, blue and grey, respectively. Figure 2e and 2f plot $g_{LL}\nu_{LL}$ versus Landau level counting N for the hole- and electron-like sequences. After normalization by the degeneracy ($g_{LL} = 6$ for the hole side and $g_{LL} = 2$ for the electron side), $\nu_{LL}$ become half-integers with half-integer intercepts at zero carrier density. This half-integer quantization is consistent with Dirac band crossings at $\nu = -2$. The reduced electron-side degeneracy likely reflects band-structure asymmetry and velocity renormalization in the moiré bands (Fig.1b)

The observed half-integer Landau fan sequence, with six-fold degeneracy on the hole side and two-fold degeneracy on the electron side at $|E| = 0.0\ V\cdot nm^{-1}$, persists over a wide range of electric fields ($|E| < 0.3\ V\cdot nm^{-1}$; Extended Data Fig. 4). This indicates that the Landau level quantization and associated degeneracies are robust over an extended electric-field range, consistent with a stable Dirac semimetal band structure near $\nu = -2$.

**Temperature dependence of the Dirac semimetal**

We further examine the temperature dependence of the Dirac semimetal phase. Figures 3a and 3b show the longitudinal resistance $R_{xx}$ as a function of moiré filling factor $\nu$ in the 3.65° twisted $WSe_2$ device at selected temperatures from 1.4 to 24 K (see extended Data Fig.5 for a second device with similar results). Measurements were performed at $E = 0\ V\cdot nm^{-1}$ (Fig. 3a) and $E = -0.5\ V\cdot nm^{-1}$ (Fig. 3b). At $E = 0\ V\cdot nm^{-1}$, $R_{xx}$ at $\nu = -2$ decreases slightly upon cooling, while fillings away from $v = -2$ show good metallic temperature dependences. By contrast, at $E = -0.5\ V \cdot nm^{-1}$, $R_{xx}$ at $v = -2$ rises sharply upon cooling.

Figure 3c summarizes $R_{xx}$ from 1.4 to 30 K for three representative cases: $v = -2$ at $E = 0\ V \cdot nm^{-1}$ (black), $v = -2.5$ at $E = 0\ V \cdot nm^{-1}$ (red), and $v = -2$ at $E = -0.5\ V \cdot nm^{-1}$ (blue) (see Extended Data Fig. 6 for the continuous electric-field dependence of $R_{xx}$ at varying temperatures). At $v = -2$ at $E = -0.5\ V \cdot nm^{-1}$, $R_{xx}$ increases nearly an order of magnitude upon cooling and saturates near $\frac{h}{2e^2}$ at the base temperature, consistent with a quantum spin Hall insulating insulator (Figure 1d - f). At $v = -2.5$ at $E = 0\ V \cdot nm^{-1}$, $R_{xx}$ decreases from ~1 kΩ to ~100 Ω, indicative of a metallic state. At $v = -2$ at $E = 0\ V \cdot nm^{-1}$, $R_{xx}$ decreases only slightly from ~1.8 kΩ to ~1.6 kΩ, reflecting weak metallic behavior, consistent with a Dirac semimetal.

**Twist angle evolution of the Quantum spin Hall and Dirac phases**

Finally, we investigate the evolution of the QSH and Dirac semimetal phases under varying twist angles. The left panels of Figures 4a-g show $R_{xx}$ as a function of moiré filling factor $v$ and vertical electric field $E$ in twisted $WSe_2$ devices with $\theta$ ranging from $\theta = 2.29°$ to $\theta = 3.89°$. The right panels show $R_{xx}$ as a function of $v$ and out-of-plane magnetic field $B_{\perp}$ measured at $E = 0\ V \cdot nm^{-1}$ in the same devices. For twist angles below 3° (Figs. 4a-b), $R_{xx}$ remains plateaued at $\sim \frac{h}{2e^2}$ across $E = 0\ V \cdot nm^{-1}$ (see extended Data Fig. 7 for linecuts along electric fields at fixed $v = -2$), consistent with well-developed QSH phases as reported previously. For these twist angles, no clear Landau fan develops up to the highest magnetic fields of 14 T, indicating low carrier mobility characteristic of moiré flat bands.

In contrast, for twist angles above 3.3° (Fig. 4c-g), $R_{xx}$ exhibits a resistance dip near zero electric fields, while a plateau or peak near $\sim \frac{h}{2e^2}$ develops only after $E$ reaches a critical value $E_c$ (black arrows). In these same devices, at near-zero electric field, well-defined Landau fans emerge at relatively low magnetic fields (~ 4T), indicating enhanced carrier mobility. The Landau level filling sequence follows that observed in Fig. 2, consistent with the six-fold degenerate Dirac fermions robust against localization(Ref. [47]). $E_c$ therefore marks the phase boundary between the QSH phase and Dirac semimetal phase.

Figure 5 summarizes $E_c$ as a function of twist angles $\theta$. For $\theta < 3.3°$, $E_c = 0$, and only the QSH phase is present. For $\theta > 3.3°$, $E_c$ is finite and increases approximately linearly with $\theta$; below $E_c$, the Dirac semimetal phase is observed, while for $|E| > E_c$, the QSH phase is stabilized. Linear extrapolation of $E_c$ versus $\theta$ reveals a critical twist angle of $\theta_c \approx 3.3°$, below which the system hots only the QSH phases.

**Robustness of Dirac Phase**

Finally, we provide a qualitative understanding of the Dirac semimetal phases observed at extended twist angles $\theta > \theta_c$ and electric fields $| E | < E_c$. On the hole side, the Landau fan sequence is consistent with a six-fold degenerate Dirac point at $\nu = -2$. While a Dirac semimetal is generically expected at a critical point separating a quantum spin Hall (QSH) insulator from a trivial insulator, the emergence of an extended Dirac semimetallic phase over a finite parameter range requires a protecting symmetry in momentum space. In contrast to graphene-based systems, where $C_{2z}T$ protects Dirac points by quantizing the Berry phase to π[48,49], the Dirac points in twisted $WSe_2$ are likely protected by $C_{2x}T$. Here $C_{2x}$ exchanges layers and valleys, while $T$ reverses valleys. Each valley hosts two inequivalent $C_{2x}T$-invariant lines, along which Dirac points are locally stable and can move without gapping unless annihilated. The observed six-fold degeneracy arises from $C_3$ rotational copies within a valley together with symmetry-related points in the opposite valley. On the electron side, the reduced degeneracy is likely due to spontaneous $C_3$ symmetry breaking upon electron doping, leading to lifting and splitting of the $C_3$-related Dirac cones and resulting in an effective two-fold valley degeneracy. This scenario also naturally explains the pronounced bending of the electron-side Landau fan, indicative of additional carriers and strong asymmetry.

## Conclusion

In summary, we report six-fold degenerate Dirac fermions in an Ising spin–orbit-coupled system emerging across a quantum spin Hall (QSH) phase transition in twisted $WSe_2$. In a 3.65° device, a QSH phase appears at high electric fields with nearly quantized resistance $h/(2e^2)$, while a Dirac semimetal phase emerges near zero field. Magnetotransport measurements reveal a half-integer Landau fan sequence with a π Berry phase, exhibiting six-fold degeneracy on the hole-doped side and two-fold degeneracy on the electron-doped side, with particle–hole asymmetry reflecting Fermi-velocity anisotropy among Dirac cones. Systematic variation of twist angle and electric field establishes a complete phase diagram of the QSH and Dirac semimetal phases. These results demonstrate that Dirac fermions can survive strong Ising spin–orbit coupling, and that the Dirac gap closure extends over a finite range of twist angle and electric field, revealing an extended topological phase transition likely enabled by broken inversion symmetry and stabilized by spin–orbit coupling and symmetry-constrained band hybridization. These findings establish twisted transition metal dichalcogenides as a versatile platform for spin-polarized Dirac quasiparticles and the interplay between Dirac physics, spin–orbit coupling, and topology, with broad implications for spintronic and topological quantum materials.

## Methods

### Device fabrication

Quad-gate $tWSe_2$ devices were fabricated using the tear-and-stack and layer-by-layer dry transfer technique[50]. Briefly, monolayer $WSe_2$, hexagonal boron nitride (hBN), and few-layer graphite flakes were mechanically exfoliated from their bulk crystals onto silicon wafers with a 285 nm $SiO_2$ layer. The monolayer $WSe_2$ flakes were bisected using an AFM probe. Subsequently, these flakes were sequentially picked up using a stamp composed of a polycarbonate (PC) film, and polydimethylsiloxane (PDMS), in the following order: the contact-gate hBN, a portion of the $WSe_2$ monolayer, the remaining $WSe_2$ monolayer with a small twist angle, the bottom-gate hBN, and the bottom-gate graphite. The completed stack was then released onto a Si/$SiO_2$ substrate pre-patterned with Ti/Pt (2/23 nm) electrodes in a Hall bar geometry, at a temperature of 180 °C. The contact gate patterns were defined on the contact-gate hBN using standard electron-beam lithography, followed by electron-beam evaporation of Ti/Pd (2/13 nm) to form the contact gates, thereby reducing the contact resistance. Finally, the top-gate graphite and top-gate hBN were transferred onto the contact gates to complete the device. The underlying Si substrate served as a depletion gate to deactivate the regions of $tWSe_2$ covered exclusively by the top gate, ensuring they did not contribute to the transport.

### Electrical measurements

The electrical measurements were performed in a closed cycle $^4$He cryostat (Cryogenic) equipped with a 18T superconducting magnet. Low-frequency (<23 Hz) lock-in techniques were used to measure the sample resistance under a small bias current (<10 nA) to minimize sample heating. Both the source-drain current and the voltage drop at the probe electrode pairs were simultaneously recorded. For the local transport measurements, the device was biased symmetrically along the major axis of the Hall bar to ensure a uniform current flow (see inset of Fig. 1d). For the nonlocal measurements, the current was injected perpendicular to the major axis (inset of Fig. 1e), forming an effective H-bar geometry.

## Figures

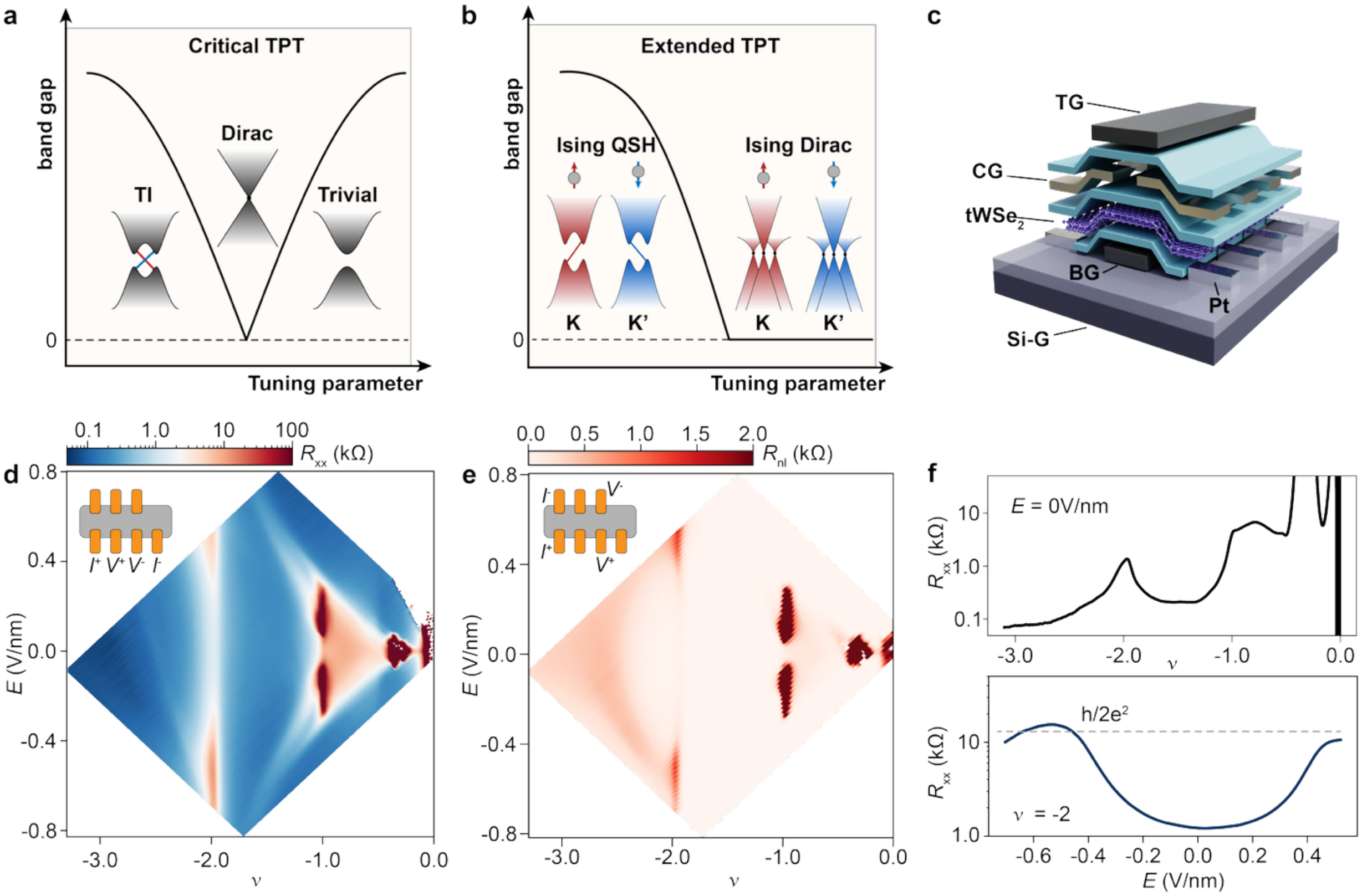

**Figure 1| Ising Dirac fermion across a Topological phase transition (TPT). a, b,** Schematic illustration of a critical topological phase transition (critical TPT) and an extended topological phase transition (extended TPT). In a, the band gap closes and reopens as material parameters are tuned, describing a transition from a topological insulator (TI) to a trivial insulator; at the critical point, the gap closes and a Dirac node forms. In b, the band gap closes without reopening upon continuous tuning of a material parameter, leading to a transition from an Ising-type quantum spin Hall (QSH) insulator (spin up and down indicated by red and blue arrows, respectively) to an Ising-type Dirac semimetal with six Dirac nodes (three per valley/spin). **c**, Device structure of a quad-gate twisted $WSe_2$ device. The top gate (TG) and bottom gate (BG) control carrier density and vertical displacement field in twisted $WSe_2$. A platinum contact gate (CG) is used to heavily dope the interface between $WSe_2$ and the platinum electrodes. A silicon back gate (Si-G) suppresses residual carriers outside the measurement channel. **d**, **e** Longitudinal resistance ($R_{xx}$) and nonlocal resistance ($R_{nl}$) as functions of moiré filling factor $v$ and vertical electric field $E$. Insets show the corresponding measurement configurations. **f**, Upper panel: $R_{xx}$ as a function of $v$ at fixed electric field $E = 0$ V/nm. Lower panel: electric field dependence of $R_{xx}$ at $v = -2$. The grey dashed line marks the quantum resistance $\frac{h}{2e^2}$. All measurements are performed at 1.5 K.

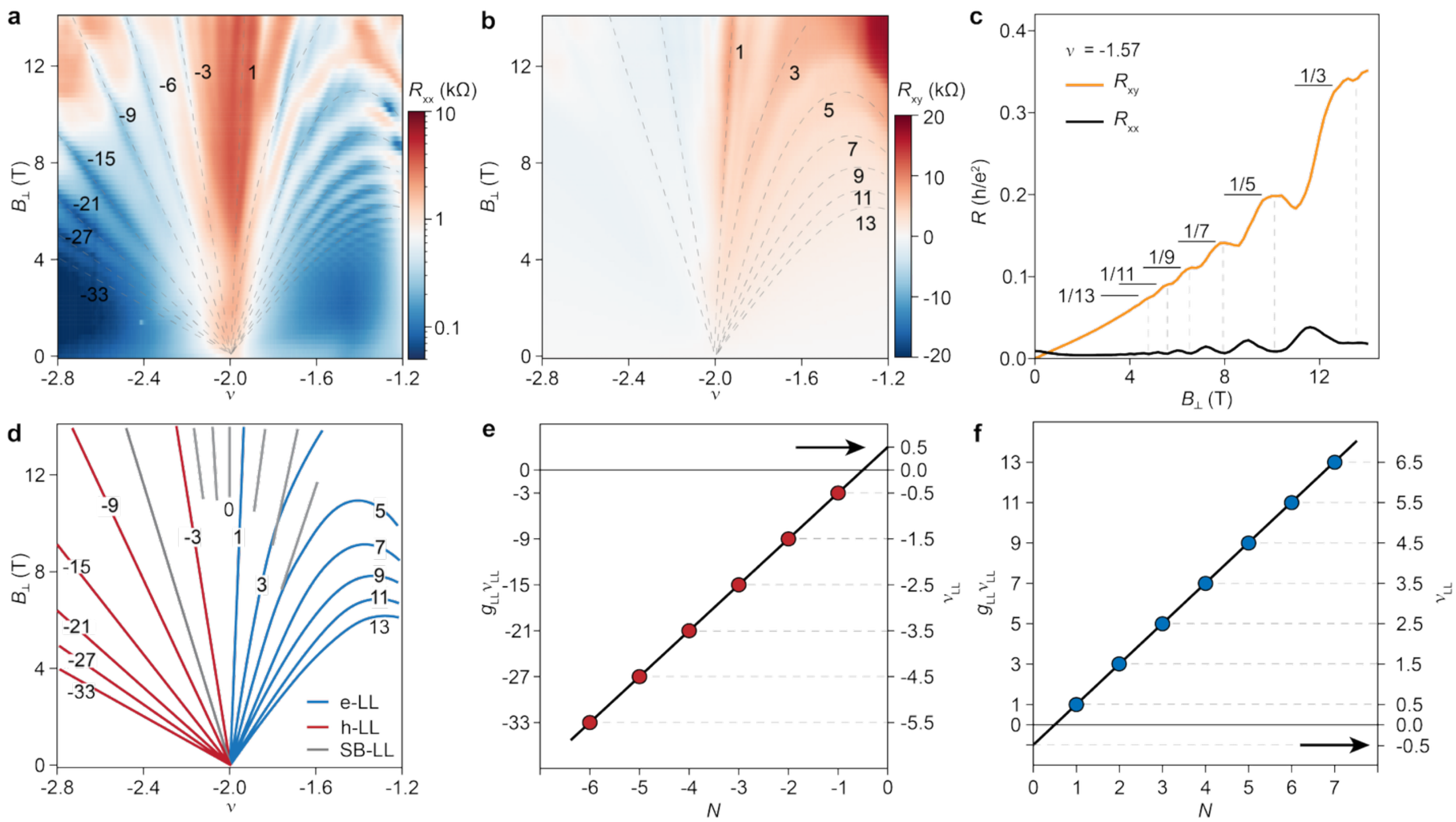

**Figure 2| Landan Fan diagram at $E = 0$ V/nm in a 3.65° twisted WSe₂ device . a, b,** Symmetrized longitudinal resistance $R_{xx} = \frac{R_{xx}(B_\perp)+R_{xx}(-B_\perp)}{2}$ (**a**) and antisymmetrized Hall resistance $R_{xy} = \frac{R_{xy}(B_\perp)-R_{xy}(-B_\perp)}{2}$ (**b**) as functions of moiré filling factor $\nu$ and out-of-plane magnetic field $B_\perp$. Dashed lines mark the minima in $R_{xx}$ and corresponding Hall plateaus in $R_{xy}$. In **a**, for $R_{xx}$ dips whose positions vary linearly with $B_\perp$, with slopes determined by the Streda formula: $\frac{\partial \nu}{\partial B_\perp} = \frac{g_{LL}\nu_{LL}}{n_M}\frac{e}{h}$, where $g_{LL}$is the Landau level degeneracy, $\nu_{LL}$ is the Landau level filling factor, $n_M$ is the moiré density, h is the Planck's constant and e is the elementary charge. **c,** Magnetic field dependence of $R_{xx}$ (black) and $R_{xy}$(orange) at fixed moiré filling factor $\nu = -1.57$. Black bars indicate the quantized resistance values $\frac{h}{Ne^2}$ with $N$ = 3, 5, 7, 9, 11 and 13. **d,** Extracted Landau fan diagram as a function of $\nu$ and $B_\perp$. Red linear trajectories correspond to six-fold degenerate hole-like Landau levels (h-LL) with $g_{LL}\nu_{LL}$ ranging from -3 to -33. Blue arc-shaped trajectories correspond to two-fold degenerate electron-like Landan levels (e-LLs) with $g_{LL}\nu_{LL}$ ranging from 1 to 13. Grey linear trajectories denote symmetry-breaking Landau levels (SB-LLs) with reduced degeneracies. **e, f,** $g_{LL}\nu_{LL}$ as a function of Landau level counting number N measured from charge neutrality for hole-like (**e**) and electron-like (**f**) sequences. The right axes show the $\nu_{LL}$. All measurements were performed at $T$ = 1.5 K.

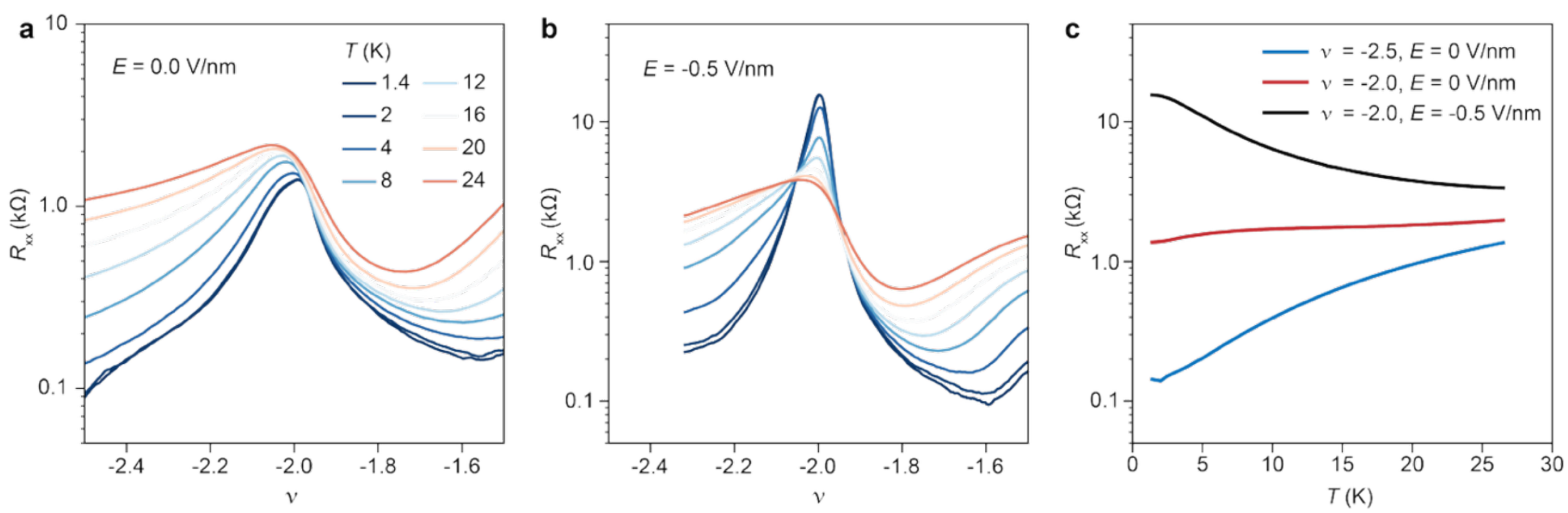

**Figure 3| Temperature dependence of the Dirac semimetal in a 3.65° twisted $WSe_2$ device. a**, **b**, Longitudinal resistance $R_{xx}$ as a function of moiré filling factor $v$ at selected temperatures from 1.4 to 24K, measured at $E = 0.0\ V \cdot nm^{-1}$ (**a**) and -0.5 $V \cdot nm^{-1}$ (**b**), respectively. **c,** Temperature dependence of $R_{xx}$ at selected $v$ and $E$:$v = -2.5;\ E = 0\ V \cdot nm^{-1}$ (blue), $v = -2, E = 0V/nm$ (red) and $v = -2, E = -0.5V/nm$ (black). All electrical measurements were performed at $B_{\perp} =$ 0 T.

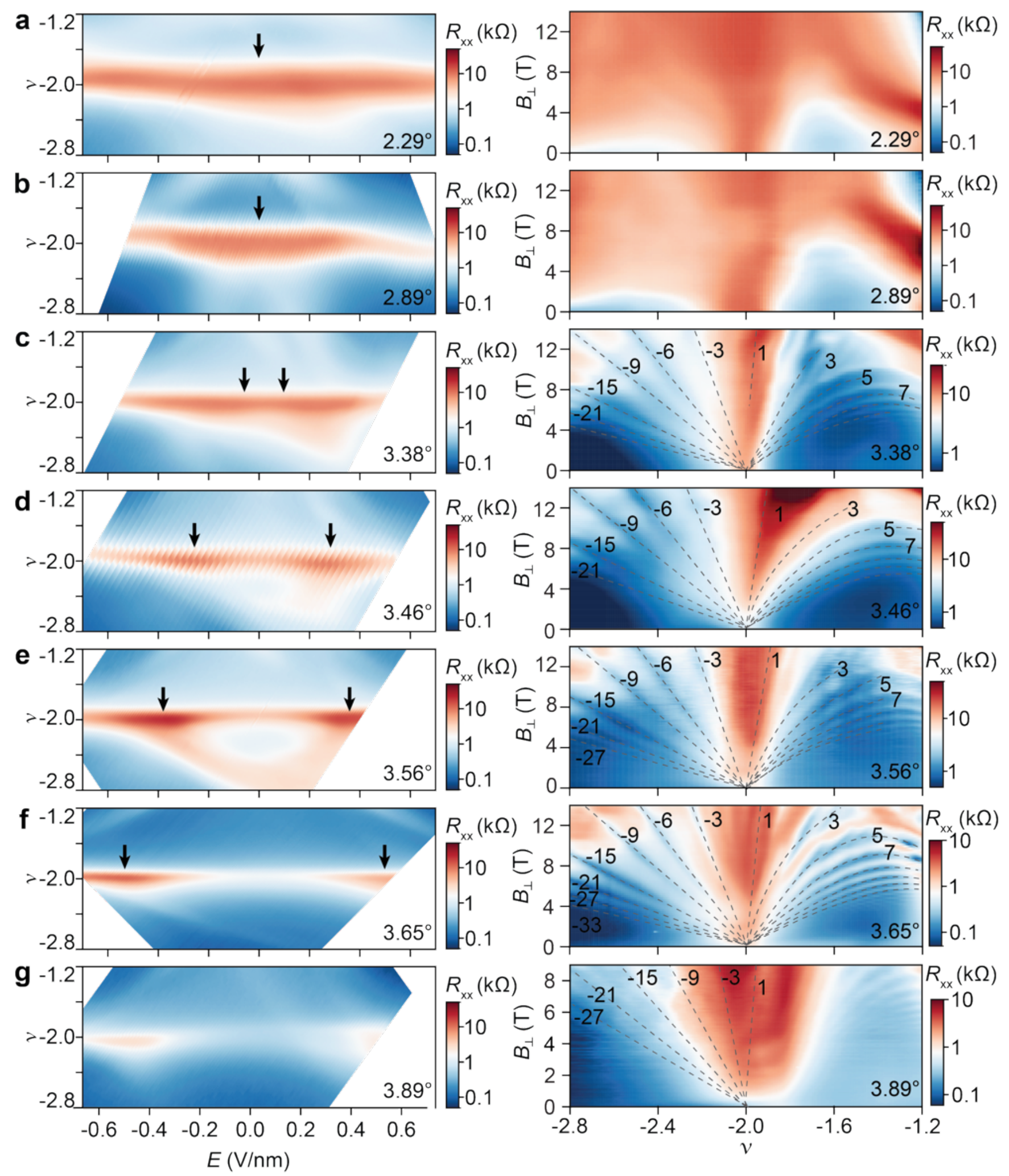

**Figure 4 | Twist-angle dependence of the quantum spin Hall and Dirac phases. a-g**, Left panels: Longitudinal resistance $R_{xx}$ as a function of moiré filling factor v and electric displacement field $E$ at $B_\perp = 0$ T, measured in twisted $WSe_2$ devices with twist angles ranging from 2.29° to 3.89°. Black arrows indicate the critical displacement fields $E_c$, defined by the peak or saturation of $R_{xx}$ near $\nu = -2$. Right-panels: $R_{xx}$ as a function of v and out-of-plane magnetic field $B_\perp$ at $E = 0$ V/nm for the same devices shown in a. Dashed lines in the lower five panels trace the Landau level sequences associated with emergent Dirac fermions near $\nu = -2$. All measurements were performed at $T = 1.5$ K.

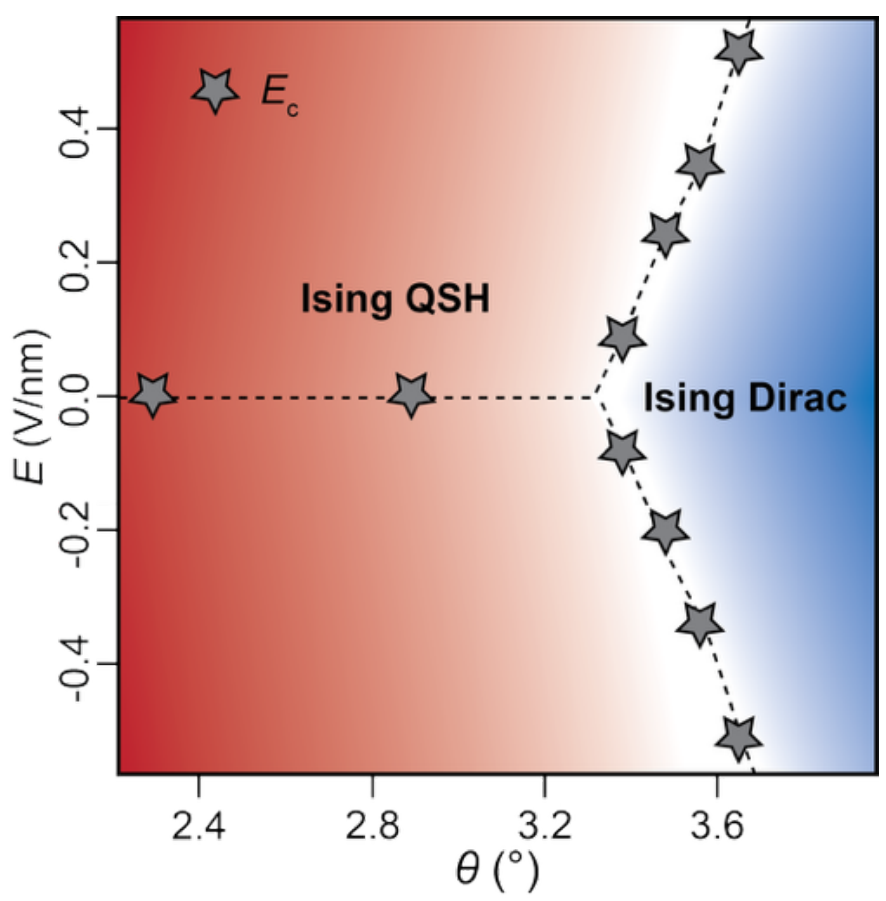

**Figure 5 | Phase diagram of the quantum spin Hall and Dirac phases.** Grey stars denote the critical displacement fields $E_c$ extracted from the measurements in Fig. 4, plotted as a function of twist angle. The red shaded region corresponds to the Ising quantum spin Hall (QSH) phase, while the blue shaded region corresponds to the Ising Dirac phase. The phase boundary is determined by the critical displacement field $E_c$, which separates the insulating QSH state from the gapless Ising Dirac state.

**Extended Data Figures**

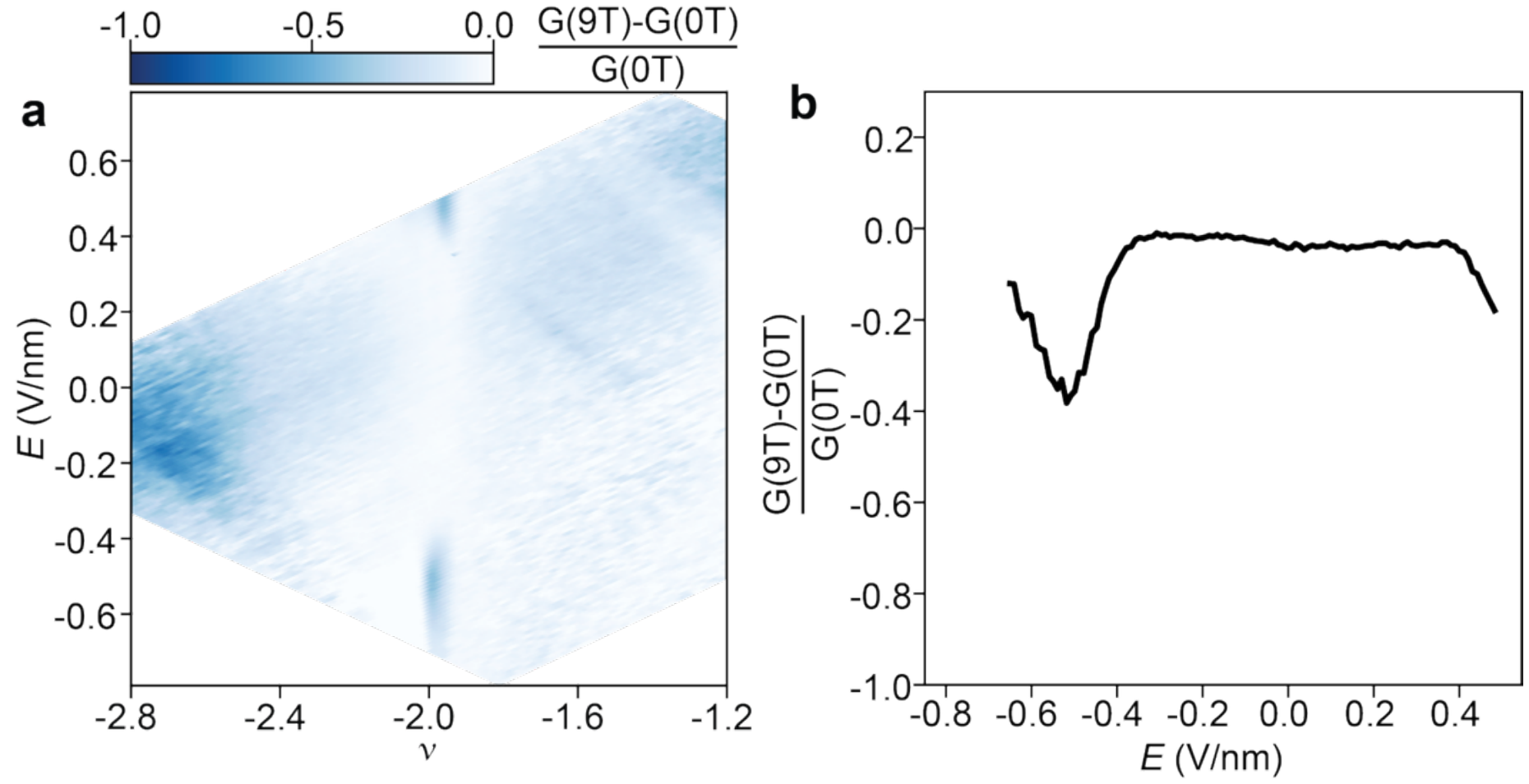

**Extended Data Fig. 1. Negative in-plane magnetoconductance in the quantum spin Hall phase. a,** Inplane magnetoconductance $\frac{\mathrm{G(9\,T)-G(0\,T)}}{\mathrm{G(0\,T)}}$ as functions of $v$ and $E$. b, electric field $E$ dependence of inplane magnetoconductance $\frac{\mathrm{G(9\,T)-G(0\,T)}}{\mathrm{G(0\,T)}}$ at $v = -2$. For $|E| \leq 0.3\ V \cdot nm^{-1}$, $\frac{\mathrm{G(9\,T)-G(0\,T)}}{\mathrm{G(0\,T)}}$ is nearly zero, indicating bulk-dominated transport consistent with the Dirac semimetal phase. For $|E| > 0.3$ V/nm, pronounced negative $\frac{\mathrm{G(9\,T)-G(0\,T)}}{\mathrm{G(0\,T)}}$ is observed, indicating a helical-edge-dominated transport characteristic of the quantum spin Hall phase.

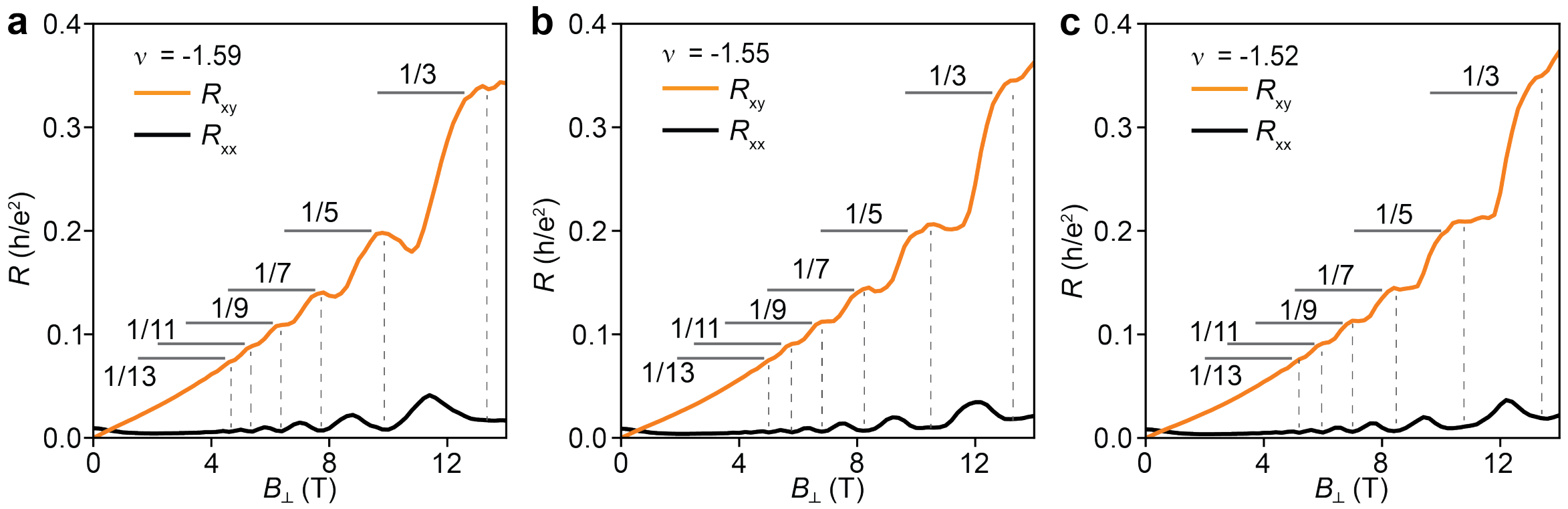

**Extended Data Fig. 2. Electron-like Landau levels at other moiré filling factors. a-c,** Longitudinal resistance $R_{xx}$ (black) and Hall resistance $R_{xy}$ (orange) as a function of out-of-plane magnetic field $B_{\perp}$ at selected moiré filling factors $v = -1.59$, -1.55 and -1.52. Grey bars indicate the quantized resistance values of $\frac{h}{Ne^2}$, with N taking odd integer values from 3 to 13. For all selected moiré filling factors, well-developed $R_{xy}$ plateaus are observed at quantized values $\frac{h}{3e^2}$, $\frac{h}{5e^2}$, $\frac{h}{7e^2}$, $\frac{h}{9e^2}$, $\frac{h}{11e^2}$ and $\frac{h}{13e^2}$, accompanied by vanishing $R_{xx}$ ($< 0.05$h/e$^2$). The quantized $R_{xy}$ and suppressed $R_{xx}$ are consistent with two-fold-degenerate Landau level sequence of $g_{LL}\ v_{LL} =$ 3, 5, 7, 9, 11 and 13. After normalization by the degeneracy factor of 2, the effective Landau filling indices $v_{LL}$ follow a half-integer sequence 1.5, 2.5, 3.5, 4.5, 5.5 and 6.5.

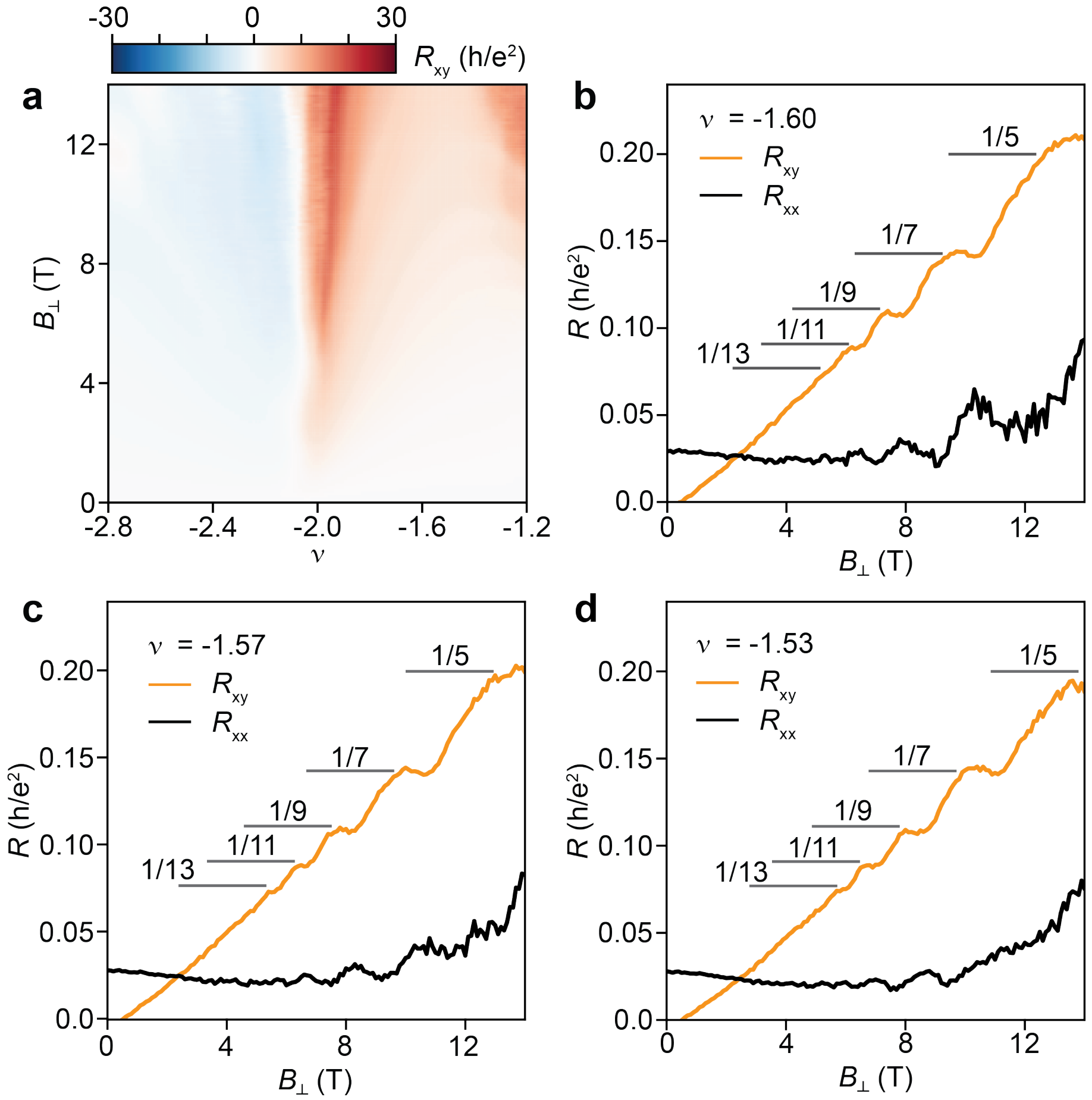

**Extended Data Fig. 3. Electron-like Landau levels with odd-integer quantization in an additional device with twist angle 3.56°. a**, Hall resistance $R_{xy}$ as a function of moiré filling factor $v$ and out-of-plane magnetic fields $B_{\perp}$ measured at $E = 0\ V \cdot nm^{-1}$. **b-d,** Longitudinal resistance $R_{xx}$ (black) and Hall resistance $R_{xy}$ (orange) a function of $B_{\perp}$ at selected moiré filling factors of $v = -1.60$, -1.57 and -1.53. Grey bars indicate the quantized resistance values $\frac{h}{Ne^2}$, with N = 5, 7, 9, 11 and 13. Well-developed $R_{xy}$ plateaus are observed with vanishing $R_{xx}$, indicating the formation of a Landau fan sequence $g_{LL}v_{LL} = $ = 5, 7, 9, 11, 13, consistent with a 2-fold degenerate Dirac fermion sequence.

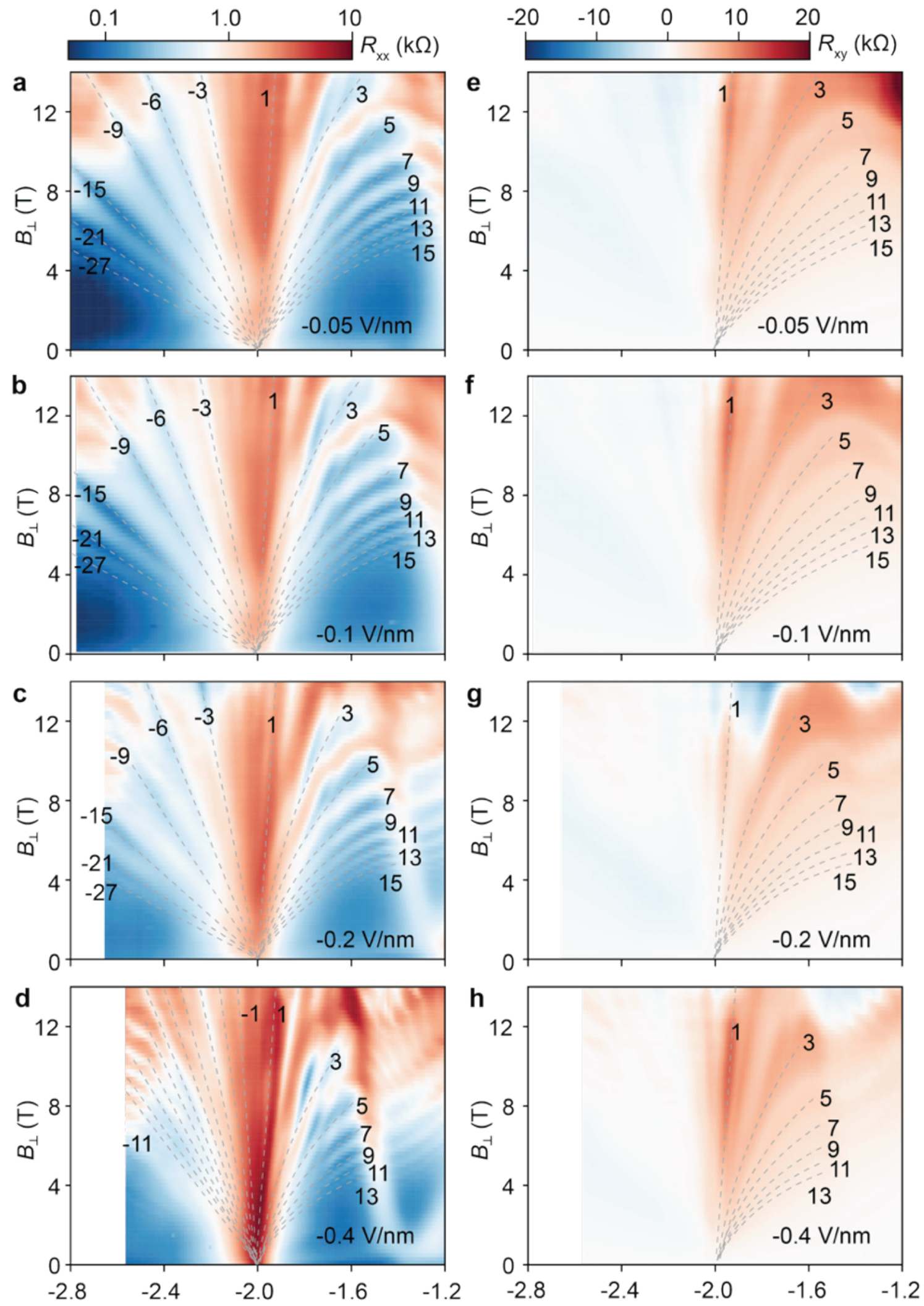

**Extended Data Fig. 4. Electric-field evolution of the Landau fan diagram in the Ising Dirac phase. a-d,** Longitudinal resistance $R_{xx}$ as a function of moiré filling factor $v$ and $B_{\perp}$ at selected electric fields $E = -0.05, -0.1$, -0.2 and -0.4 $V \cdot nm^{-1}$. Grey dashed lines trace the Dirac Landau fan sequences labeled by their corresponding $g_{LL}v_{LL}$. **e-h,** Hall resistance $R_{xy}$ as functions of $v$ and $B_{\perp}$ at the same electric fields as in **a-d**. Grey dashed lines follow the quantized Hall plateaus, with the corresponding Landau level sequence numbers indicated for each trajectory.

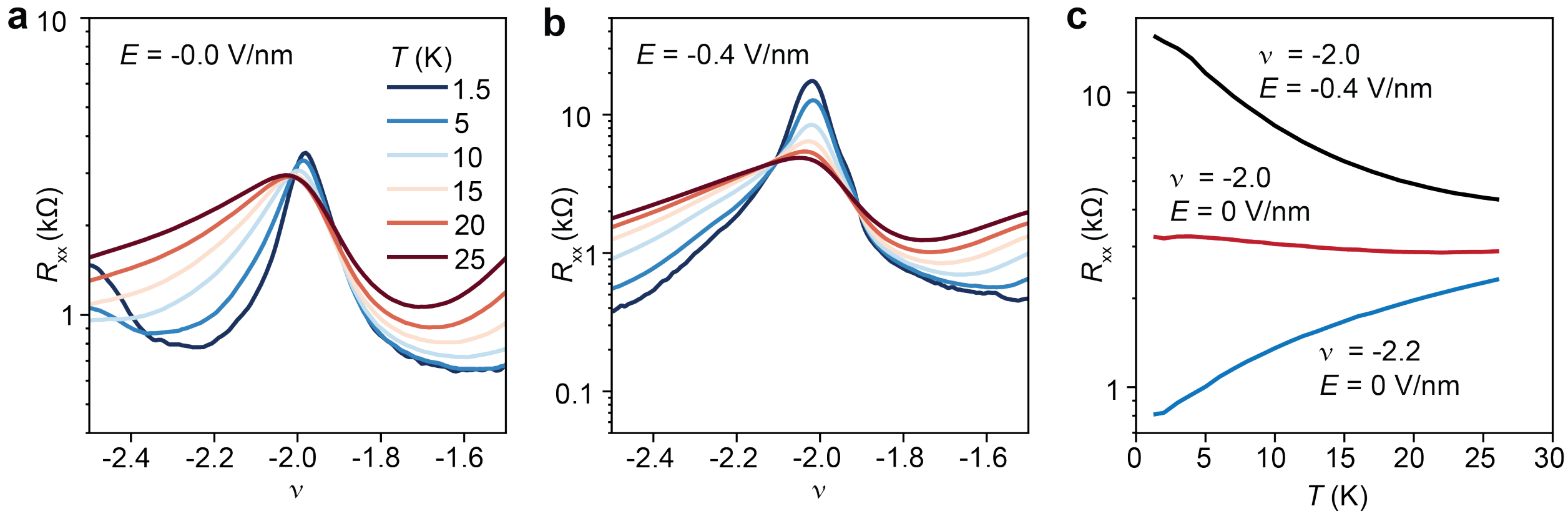

**Extended Data Fig. 5. Temperature dependence of the Dirac semimetal in an additional device. a, b,** Longitudinal resistance $R_{xx}$ as a function of $v$ measured at $E = 0\ V \cdot nm^{-1}$ (**a**) and $E = -0.4\ V \cdot nm^{-1}$ (**b**) for selected temperatures from 1.5 to 25K. Curves with the same color in the legend correspond to the same temperature. **c,** Temperature dependence of $R_{xx}$ at selected moiré filling factor $v$ and electric field $E$: $v = -2.0, E = -0.4V \cdot nm^{-1}$ (black, QSH phase), $v = -2.0, E = 0V \cdot nm^{-1}$ (red, Dirac semimetal phase) and $v = -2.2, E = 0\ V \cdot nm^{-1}$ (blue, normal metal phase). The Dirac semimetal phase shows weak temperature dependence, whereas $R_{xx}$ increases by an order of magnitude in the QSH phase (black) and decreases by nearly an order of magnitude in the normal metal phase (blue).

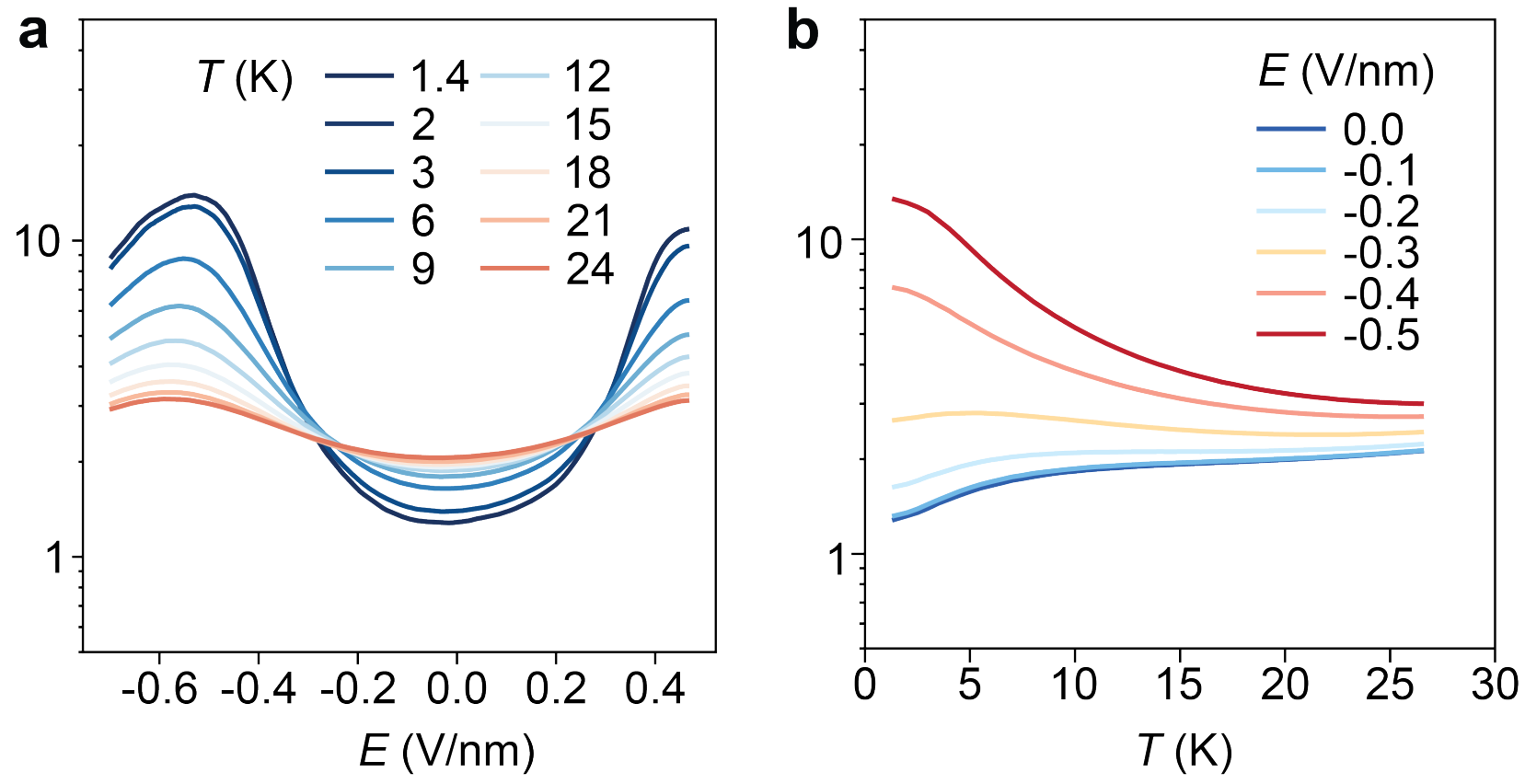

**Extended Data Fig. 6. Electric field dependence of $R_{xx}$ at varying temperatures. a,** Electric field dependence of $R_{xx}$ at $v = -2$ and selected temperatures ranging from 1.4K to 24K. **b**, Temperature dependence of $R_{xx}$ at selected electric fields $E$ = 0.0, -0.1, -0.2, -0.3, -0.4 and -0.5 $V \cdot nm^{-1}$. For $|E| \leq 0.3\ V \cdot nm^{-1}$, $R_{xx}$ weakly decreases with decreasing temperature, consistent with semimetallic transport behavior. For $|E| > 0.3\ V \cdot nm^{-1}$, $R_{xx}$ increases upon cooling, and saturates below 20 kΩ, consistent with a quantum spin Hall phase characterized by an insulating bulk gap and conducting helical edge states.

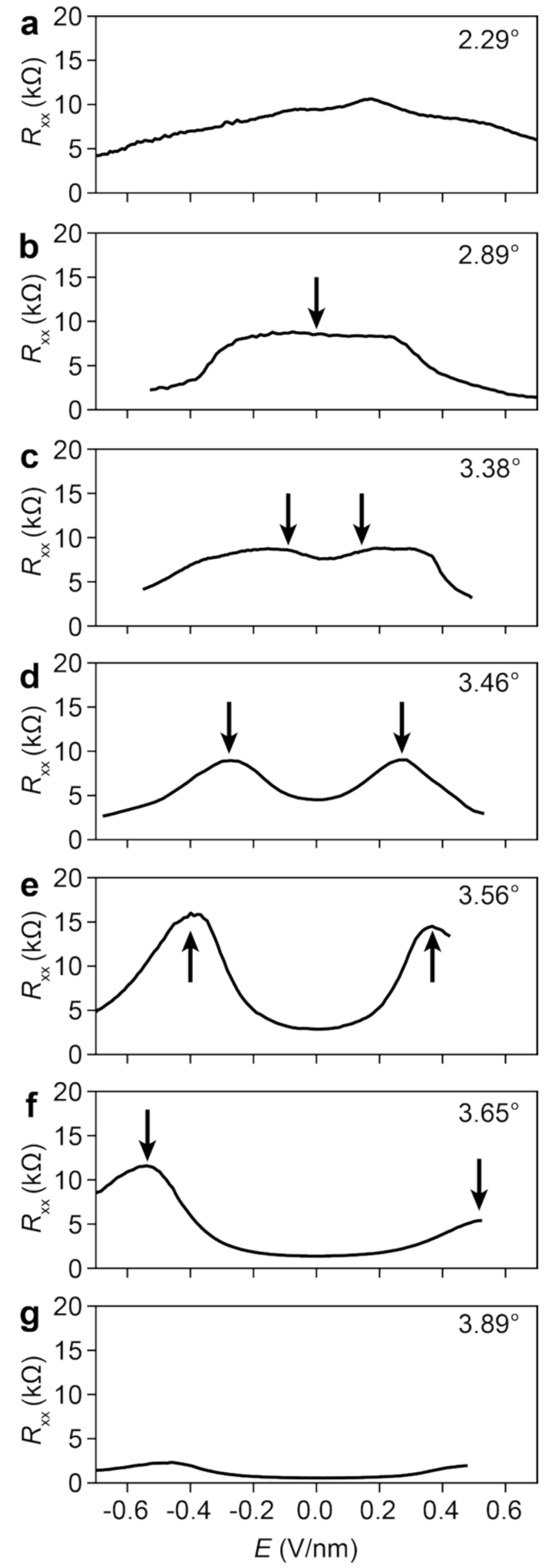

**Extended Data Fig. 7. Electric-field dependence of $R_{xx}$ at $v = -2$ for different twist angles. a-g,** Longitudinal resistance $R_{xx}$ as a function of electric fields $E$ at $v = -2$ for twist angles ranging from 2.29° to 3.89°. Black arrows indicate the critical electric fields $E_c$ where $R_{xx}$ peaks or forms a plateau. For twist angles < 3.3°, the $R_{xx}$ plateau is continuous across $E = 0$ V/nm. For twist angles above 3.3°, a resistance dip emerges for $|E| < E_c$, corresponding to the Ising Dirac phase.